# On the use of multilayer Laue lenses with X-ray Free Electron Lasers

**Mauro Prasciolu,[a] Kevin T. Murray,[a] Nikolay Ivanov,[a] Holger Fleckenstein,[a] Martin Domaracký,[a] Luca Gelisio,[a] Fabian Trost,[a] Kartik Ayyer,[b] Dietrich Krebs,[a,c] Steve Aplin,[d] Salah Awel,[a] Ulrike Boesenberg,[d] Anton Barty,[a] Armando D. Estillore,[a] Matthias Fuchs,[e] Yaroslav Gevorkov,[a,f] Joerg Hallmann,[d] Chan Kim,[d] Juraj Knoška,[a,c] Jochen Küpper,[a,c,g] Chufeng Li,[a] Wei Lu,[d] Valerio Mariani,[a] Andrew J. Morgan,[h] Johannes Möller,[d] Anders Madsen,[d] Dominik Oberthür,[a] Gisel E. Peña Murillo,[a,c] David A. Reis,[i] Markus Scholz,[d] Božidar Šarler,[j,k] Pablo Villanueva-Perez,[l] Oleksandr Yefanov,[a] Kara A. Zielinski,[a] Alexey Zozulya,[d] Henry N. Chapman,[a,c,g] and Saša Bajt[a,g,*]**

[a]Center for Free-Electron Laser Science CFEL, Deutsches Elektronen-Synchrotron DESY, Hamburg, Germany
[b]Max Planck Institute for the Structure and Dynamics of Matter, Hamburg, Germany
[c]Department of Physics, Universität Hamburg, Hamburg, Germany
[d]European X-ray Free-Electron Laser Facility, Schenefeld, Germany
[e]University of Nebraska-Lincoln, Department of Physics and Astronomy, Lincoln, Nebraska, USA
[f]Hamburg University of Technology, Hamburg, Germany
[g]The Hamburg Centre for Ultrafast Imaging, Hamburg, Germany
[h]University of Melbourne, School of Physics, Parkville, Victoria, Australia
[i]Stanford PULSE Institute, SLAC National Accelerator Laboratory, Menlo Park, California, USA
[j]University of Ljubljana, The Faculty of Mechanical Engineering, Ljubljana, Slovenia
[k]Institute of Metals and Technology, Ljubljana, Slovenia
[l]Lund University, Synchrotron Radiation Research and NanoLund, Lund, Sweden

Multilayer Laue lenses were used for the first time to focus x-rays from an X-ray Free Electron Laser (XFEL). In an experiment, which was performed at the European XFEL, we demonstrated focusing to a spot size of a few tens of nanometers. A series of runs in which the number of pulses per train was increased from 1 to 2, 3, 4, 5, 6, 7, 10, 20 and 30 pulses per train, all with a pulse separation of 3.55 µs, was done using the same set of lenses. The increase in the number of pulses per train was accompanied with an increase of x-ray intensity (transmission) from 9% to 92% at 5 pulses per train, and then the transmission was reduced to 23.5 % when the pulses were increased further. The final working condition was 30 pulses per train and 23.5% transmission. Only at this condition we saw that the diffraction efficiency of the MLLs changed over the course of a pulse train, and this variation was reproducible from train to train. We present the procedure to align and characterize these lenses and discuss challenges working with the pulse trains from this unique x-ray source.

**Keywords**: x-rays, multilayer Laue lenses, XFEL, optics.

*Saša Bajt, E-mail: sasa.bajt@desy.de

## 1 Introduction

X-ray free electron lasers distinguish themselves from other x-ray sources with their extremely short pulses, high transverse coherence and high peak powers. These properties opened up new research areas including single shot imaging of biological samples, imaging the dynamics of matter, creating matter under extreme conditions and studying nonlinear optical processes in the hard x-ray regime. Nonlinear Compton scattering[1], for example, occurs at extreme intensities when two incoming x-ray photons of photon energy $E$ simultaneously interact with an atom to convert into a single x-ray photon that is energetically red-shifted in comparison to $2E$. The generation of a photon at precisely $2E$ corresponds to the process of second harmonic generation, which is well known in nonlinear optics and has also been observed in the x-ray regime[2]. X-ray and optical photons can also combine at high intensities and in the presence of matter in difference- and sum-frequency generation[3] and other wave-mixing processes such as parametric down conversion[4]. In general, non-linear X-ray interactions benefit from very high fluence, which can be increased by focusing the XFEL beams to a smaller spot. We wish to investigate the generation of high-intensity beams at XFEL sources using multilayer Laue lenses[5] (MLLs). This new generation of highly efficient and robust x-ray optical elements has been designed to operate under the full beam conditions of these sources[6,7]. Part of this work was done during the commissioning of the MID beamline[8,9] of the European XFEL[10] (Schenefeld, Germany) in two experimental runs

.

In general, focusing x-ray beams to a very small spot, below 10 nm, is challenging. It requires optics with little or no wavefront aberrations. Optical elements used with XFEL beams have to be also extremely robust to survive and perform reliably when exposed to high power beams. XFEL beams are typically focused with compound refractive lenses (CRLs) or reflective mirrors. To increase the numerical aperture (NA) of CRLs, which is needed to achieve very small spot size, many lenses need to be stacked behind each other[11]. This however, also reduces their efficiency. Furthermore, CRLs have strong chromatic aberrations so their ability to focus to extremely small spot sizes is limited[12]. Reflective mirrors, such as Kirkpatrick-Baez (KB) mirrors, which operate at grazing incidence, can accept the full beam. In this case the beam is spread over a large area and the intensity of the beam on the optic is low. The reflectivity and efficiency of KB mirror system can be very high. However, substrates and coatings are required that have nearly perfect surface figure and extremely low surface roughness[13]. Diffraction-based optics, such as Fresnel zone plates, can also be used to focus XFEL beams[14-17]. They also have chromatic aberrations, but their effects are generally smaller than for CRLs due to their shorter focal lengths and lower dispersion. Since these lenses consist of nanostructures printed on x-ray transparent membranes, they are fragile. Theoretical simulations predicted that rapid temperature fluctuations of many hundred degrees Kelvin lead in stress and/or strain changes followed by a failure when exposed to the direct beam[17]. However, the response of zone plates is also strongly dependent on the materials used in these nanostructures. For example, typical zone plate materials, such as gold and tungsten, cannot survive long in an intense XFEL beam. But combining high-Z metal (Ir) with excellent heat dissipation of low-Z material (diamond), as demonstrated with Ir-filled diamond zone plates, leads to increased radiation hardness[14]. Such zone plates could withstand unfocused XFEL beam at LCLS (SLAC, USA) but under pre-focused XFEL beam their damage threshold was surpassed. It was reported that the LCLS beam at 4 and 8 keV photon energies and a beam size of 2 x 2 mm$^2$ were not detrimental for grating nanostructures made of pure diamond[18]. This beam size was much larger than the typical areas of zone plates. Making zone plates this large is possible, but requires an increase in the number of zones which in turn increases its chromatic aberration, thus reducing the tolerable bandwidth.

In this paper we report about focusing XFEL beams with MLLs. These diffractive optics have much higher diffraction efficiency as compared to Fresnel zone plates and combined with high

NA could potentially focus XFEL beam to intensities required to study nonlinear processes such as mentioned above.

## 2  Multilayer Laue lenses

All MLLs used in this study were prepared in our laboratory with a custom-built magnetron sputtering system. The multilayers consisted of silicon carbide (SiC) and tungsten carbide (WC), which form sharp, smooth and stable interfaces[6,7,19]. The multilayer, which was >105 μm thick, consisted of 10048 bilayers and was deposited on a polished Si wafer substrate with initial roughness of about 0.15 nm rms. The substrate was moved between the two sputter targets in the process of spinning around its own axis while the layer thickness was controlled by varying the platter velocity.

Similar to Fresnel zone plates, the resolution of an MLL lens is determined by the finest layer. Tungsten carbide/silicon carbide (WC/SiC) multilayer systems have remarkably low interfacial roughness and can form nanometer or even sub-nanometer periods as previously demonstrated for periodic multilayers[20,21]. However, the variation in period that produces the focusing behavior in MLLs leads to stress and interface roughness variation in the deposited nanostructure. Also, the total multilayer structure is hundreds of times thicker than the multilayers usually used for reflective coatings. We overcame these challenges using new materials and by optimizing the sputtering process as reported before[7]. The refractive carbide materials used here produce multilayer structures that are thermally stable and retain their period when heated up to temperatures of 800ºC[6]. A high aspect ratio of the period of the finest layer to the thickness of the lens (in the direction of beam transmission) is needed to achieve high diffraction efficiency. An aspect ratio of several thousand is needed in x-ray regime above about 10 keV. Such high aspect ratio structures can easily be achieved when slicing an MLL from the deposited structure.

Diffraction efficiencies above 80% were achieved for energies of 17-20 keV[22] and even higher efficiencies can be reached at higher energies as we demonstrated at 60 keV[23]. To obtain high efficiency over the entire lens, it is critical that each layer is also correctly tilted to satisfy Bragg's law in each position along the lens. We developed a simple method to achieve this wedging of the layers by affixing a straight-edged mask at a pre-selected height above the substrate and spinning it together with the substrate during the deposition process[24]. A desired thickness profile is then formed in the penumbra of the mask. Because it takes a very long time to deposit such thick multilayers (~6 days to prepare 105 μm thick multilayer) we maximized the deposition output by utilizing two masks placed at different heights. Consequently, we could prepare lenses optimized for the same energy but of two different focal lengths during the same deposition run. After the deposition was finished the multilayer thickness profiles were determined by making "depth soundings" of the structure at a number of positions using a focused ion beam (FIB)[24]. This was used to identify the positions where layer spacing and layer tilt curvature matched the design[25]. MLLs were cut using a FIB. This took several days per MLL due to their large thickness. Final extraction and transfer of each MLL was performed with a nanomanipulator (Omniprobe®). Each MLL was attached to the corner of a 100 μm thick diamond wafer, welded on the side with two Pt dots (by electron-induced deposition) and finally thinned using the ion beam to an optical thickness that would give the optimal diffraction efficiency.

Experiments reported here were performed at the MID instrument[9] of European XFEL during experimental runs in April and October 2019 at 8.9 keV and 10.1 keV photon energies, respectively. MLLs optimized for these two energies were designed to focus x-rays in ideal case (unaberrated lenses) to 15 nm with focal lengths of 6 and 7 mm. Since an MLL deposited on a flat substrate focuses x-rays only in one direction, two MLLs positioned orthogonal to each other, are needed to achieve a 2D focusing. Precise alignment, including astigmatism correction[26], was achieved by mounting each lens on a separate hexapod with six degrees of freedom. Focusing was achieved with a pair of MLLs optimized for the same energy with two different focal lengths. The multilayer, from which these MLLs were prepared, was deposited starting with the thinnest and ending with the thickest layers. This took advantage of the smooth substrate. Such a structure diffracts x-rays away from the plane of the substrate. This off-axis design simplifies separating and blocking the direct beam and the two diffracted beams coming from horizontal and vertical lenses.

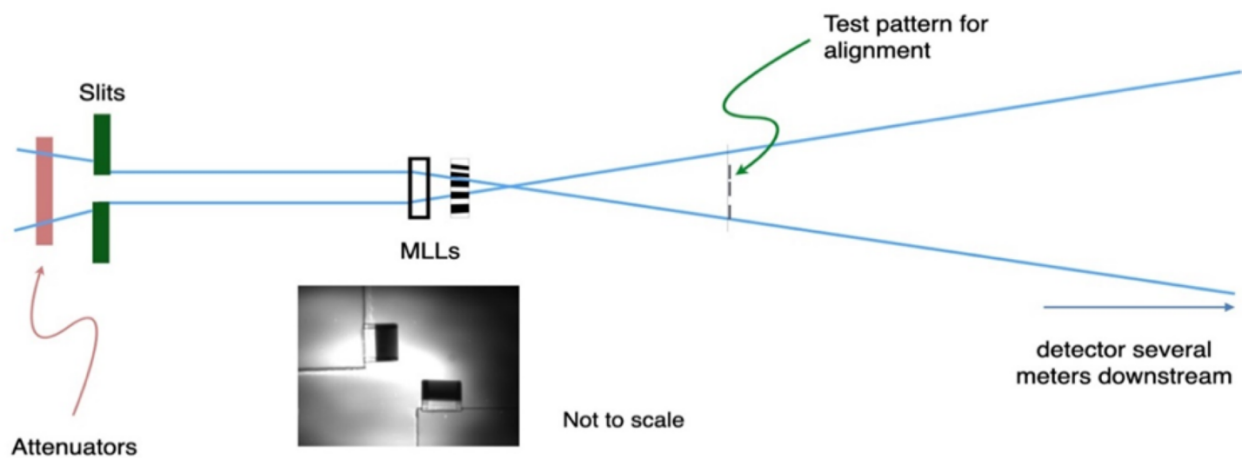

**Fig. 1** Schematic of the experimental setup showing the major components. The image below the MLLs shows horizonal and vertical focusing MLLs under Bragg condition (black) when illuminated by XFEL beam as seen with x-ray eye detector placed downstream the lenses.

*2.1 Experimental setup*

The approximate dimensions of the MLLs were 100 μm (height) x 125 μm (width) x 4.5 μm (optical thickness). Such large MLLs are needed to capture the largest possible part of highly coherent XFEL beam, without having to pre-focus it to damaging intensities. In this experimental run we worked with 8.9 keV x-rays and the beam was collimated and slightly pre-focused (~ 300 x 300 μm$^2$) using only one set of CRLs. Additional beam size reduction and removal of stray scattering and tails of the beam was accomplished with a set of slits. Nevertheless, the slitted beam over-filled the MLLs thus causing the beam to illuminate not only the lenses but also their supporting structures. Due to beam pointing instabilities and problems with data acquisition, measurements and data analysis were very difficult. In our second campaign we worked with 10.1 keV x-rays. The XFEL beam was collimated and pre-focused with two sets of CRLs. The beam illuminating MLLs was at least three times more intense as compared to experimental run in April 2019 and an additional pinhole was placed upstream from MLLs (Fig. 1). The x-ray beam path at the MID beamline was in vacuum except in the region surrounding our equipment, of about 1.2 m length. This region included an upstream pinhole, the set of two MLLs, a sample, removable x-ray eye detector and optical microscope, and a beam stop.

During the initial alignment procedure, the intensity of the XFEL beam was attenuated by several orders of magnitude. An x-ray eye detector was placed along the beam path behind the MLLs. An MLL lens in the beam path will transmit most of the incident x-rays since it is only about 4.5 µm thick. However, once it is tilted to the Bragg angle it becomes completely black because it efficiently diffracts most of the x-rays away from the optical axis. This is the case for both lenses in Fig. 1. In this image the lenses are still far apart from each other. During alignment, they are overlapped as visualized with the x-ray eye detector, to ensure that the x-rays diffracting from the first lens fall on the second, orthogonally placed lens. As an alignment sample we used a structure made out of gold using optical lithography [Fig. 2(a)]. It consisted of a 400 nm thick gold dot (8 µm diameter) surrounded by five 4 µm wide concentric rings, with increasing diameter. This structure was printed directly on a 200 µm diamond wafer and placed close to the focal plane. It was placed downstream of the MLL focus so that a highly magnified image of this structure was projected on the AGIPD (Adaptive Gain Integrating Pixel Detector)[27] detector [Fig. 2(b)]. In addition to the gold structure with concentric rings one can see several vertical lines which are liquid-air interfaces of two liquid sheet jets placed in the beam path downstream of this sample. AGIPD is high dynamic range pixel array detector with a 4.5 MHz frame rate. The projection image was used to fine tune the alignment, correct the astigmatism (observable as a different magnification of the image in the horizontal and vertical directions) and to determine the focal plane. The AGIPD was 4 m away from the interaction (focus) region. It was under vacuum and the flight path was extended in front of the detector, starting at about 80 cm from the focal plane with an entrance aperture by an off-axis 40 mm diameter, ~770 µm thick, water cooled, diamond window. In addition to the beam stop that blocked the direct and the single first order diffracted beams, a 300 µm thick Si was mounted directly on the diamond window to attenuate the diverging beam to reduce counts recorded by the AGIPD.

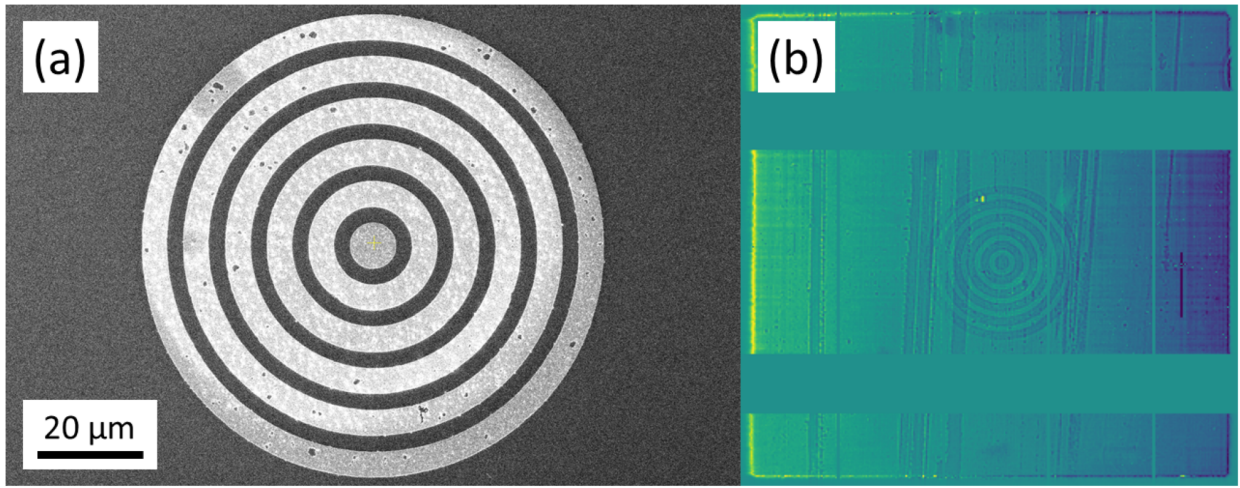

**Fig. 2** (a) An SEM image of a bull's eye (400 nm thick gold structure) printed on 200 µm diamond substrate. The structure has a diameter of 88 µm. (b) Highly magnified projection image of this same structure as seen on the AGIPD detector, which was placed 4 m downstream from the focus. The high contrast and circular shape of this structure enabled fast and precise alignment of the MLLs including astigmatism correction. The highly divergent beam spreads over 3 detector chips covering an area of about $60 \times 60$ mm$^2$. The image shows the gaps between detector modules while the sharper vertical lines are projection images of liquid sheet jets placed donwstream from the sample and focus. Imaging was performed with highly attenuated XFEL beam.

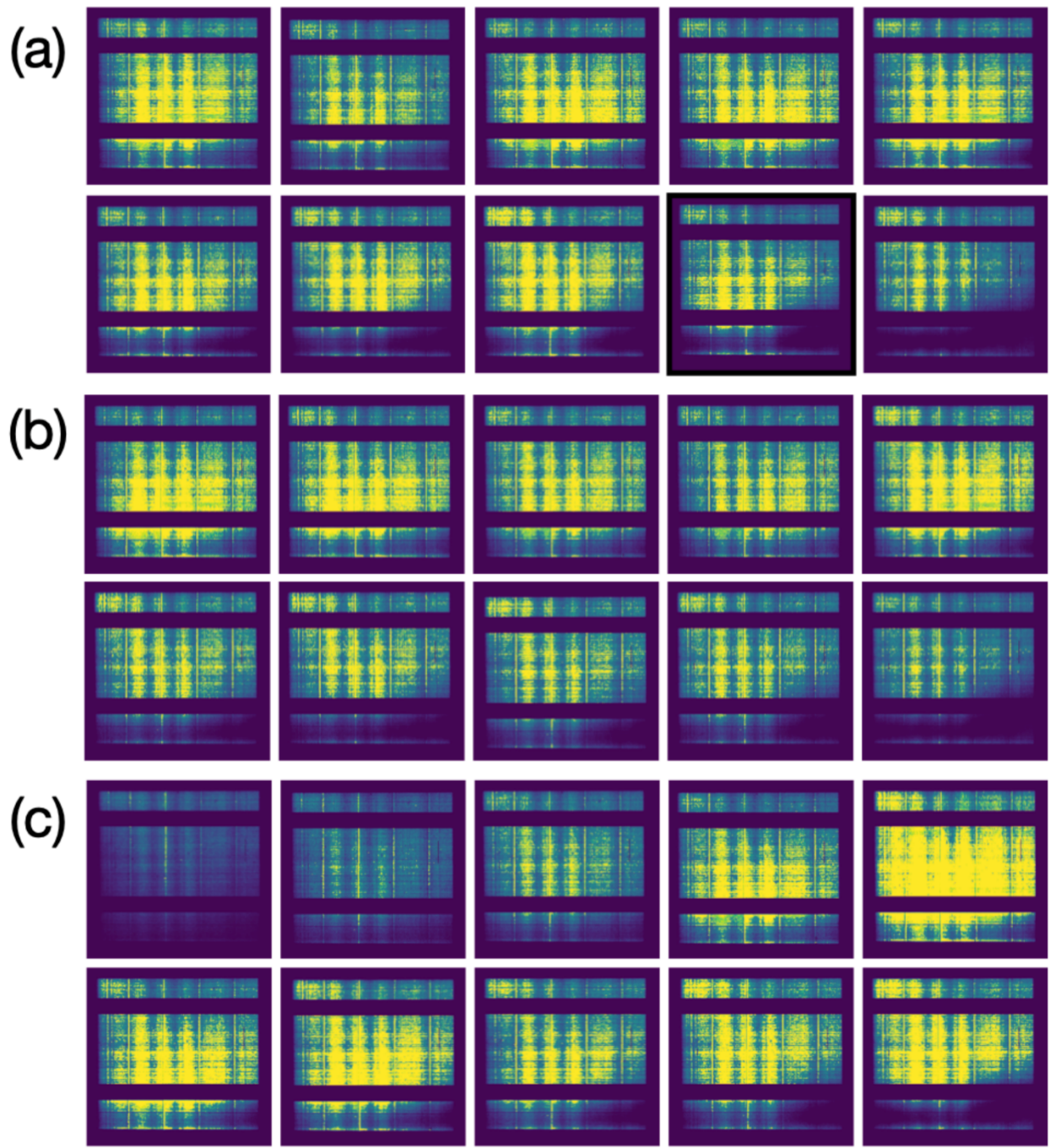

**Fig. 3** Single-pulse images of the pupil for different pulses and trains in a run. Each train consisted of 20 pulses with a separation of 3.56 μs and 23.5% transmission. The odd-numbered pulses are shown. (a) The first train in the run. (b) Train 23. (c) Train 217.

## 3  Results and discussion

*3.1 Optics characterization and results*

The wavefronts of the MLLs used in this experiment were pre-characterized using ptychographic x-ray speckle tracking (PXST). More details about the theory[28], experimental application[29] and a software code[30] for PXST can be found elsewhere. The method relies upon the ability to nicely visualize aberrations of the lens by making a projection image of a small sample placed just slightly outside the focus. This is a simple way to form magnified high-resolution images. In order to accurately map out any wavefront distortions of the beam illuminating the sample, the sample should have many small features. The sample is scanned around while magnified holographic images are collected on the detector. The image of the whole sample can then be formed by stitching together many such illuminated patches. However, if the lenses are not perfect (the wavefront not being a perfect sphere) distortions in the image are revealed when images do not perfectly overlap when these small patches are stitched together. So, when we construct the stitched image, we have to "undistort" each projected image so that they match up in the overlapping regions. This is just ptychography, but now applied in a geometric sense, rather than through diffraction. This map of the distortions represents the phase gradient, which after integration gives us the wavefront map of the lens. The focus size and distribution can be extracted once we perform a Fourier transform of the lens wavefront map and backpropagate the wavefront to the focal plane. We found that the phase of two orthogonally placed MLLs is separable[7] and the phase of each lens is similar to the other, which is to be expected since the lenses were prepared the same way. The focus size was determined using wavefront measurement and back propagating to the focal plane. Based on PXST characterization[28,29,30] the achieved focus in the focal plane was 26 nm FWHM. We estimate that the energy in focus was ~6 µJ and the pulse duration was 30 fs.

*3.2 Lens performance under pulse trains*

Here we show the performance of one set of lenses that we were testing by increasing the intensity of the incoming x-ray beam and the number of pulses per train at a pulse energy of ~1.1 mJ. There were 10 pulse trains delivered per second. After the MLLs were aligned they were used with single pulses and 92% beamline transmission for 6 hours. Apart from beam-pointing instability no problems were observed. In one study we performed a series of runs in which the number of pulses per train was increased from 1 to 2, 3, 4, 5, 6, 7, 10, 20 and 30 pulses per train, all with a pulse separation of 3.55 µs. The increase in the number of pulses per train was accompanied with an increase of x-ray intensity (transmission) from 9% to 92% at 5 pulses per train, and then the transmission was reduced to 23.5 % when the pulses were increased further. The final working condition was 30 pulses per train and 23.5% transmission. At this condition we saw that the diffraction efficiency of the MLLs changed over the course of a pulse train, and that this variation was reproducible from train to train.

Fig. 3 shows snapshots of the magnified pupil of the MLLs recorded on the AGIPD detector in a single run as a function of pulse number in the train and for three of the trains in the run with a beamline transmission of 23.5%. The two horizontal dark lines seen in each image are the gaps between detector modules on the AGIPD. The run consisted of 1000 trains, each with 20 pulses for a total of 20000 recorded frames. The time sequence goes from left to right and top to bottom. On the first pulse of each train the pupil is optimally illuminated but with increasing pulse number in the train the illumination changes. This is especially visible in the last pulse of the first train [Fig. 3(a)], which shows a reduction of intensity in the bottom right-hand side of the pupil. The observed reduction is similar to patterns observed when the lenses are not properly aligned, and we speculate that this was caused by a small rotation of the lens, induced by the x-ray pulses over the course of the train, with the lens returning to its original orientation by the start of the next train. To test this hypothesis, we changed the tilt of the lenses in such a way that we observed an improvement of the diffraction efficiency during the course of the train, as the induced motion brought the lens into alignment. The second set of images [Fig. 3(b)] was obtained from the 23$^{rd}$ train and the last set from the 217$^{th}$ train [Fig. 3(c)]. There is not much difference between the first and the second set of frames. However, the last set of frames looks quite different and is representative of the situation when the lenses were exposed to trains with many pulses. After being exposed to several trains the lenses often tended to depart from their Bragg condition at the start of the next train but become better aligned by the end of the train [Fig. 3(c)].

The final run (not shown) was done with trains with 30 pulses at 23.5% transmission. In this case the lens diffraction efficiency changed such that by the end of the 1000 trains no diffraction could be seen on the detector. Thus, we feared that both lenses were destroyed. However, this was not the case. We could verify that both lenses were still there with the x-ray eye detector but no longer in the Bragg condition. After re-aligning them we saw that they were still diffracting.

After the experiment we examined both lenses with SEM and compared them with images taken before the experiment. No obvious changes could be seen in these SEM images[30]. Later measurements of the diffraction of these lenses also did not show any change.

In another test at even higher beamline transmission during experiment 2543, where the incoming beam on the MLLs was about three times higher in intensity, we found that the upstream lens broke off its mount. One explanation for this and the wobbling of the lenses over the course of the pulse train is that the heat, caused by x-ray absorption of the lens, cannot be dissipated quickly enough. Although the lens appears to be thermally very stable the large heat gradients could cause a tilt of the lens at the points at which they are attached to the diamond mount. Another possible explanation is that the lens broke off the support due to the momentum transfer of the photons hitting the lens. Lenses were connected to the diamond substrate only via two very small Pt welding dots as indicated in Fig. 4(a). This way of mounting the MLLs was used successfully in synchrotron experiments but the conditions at XFELs are obviously much more severe.

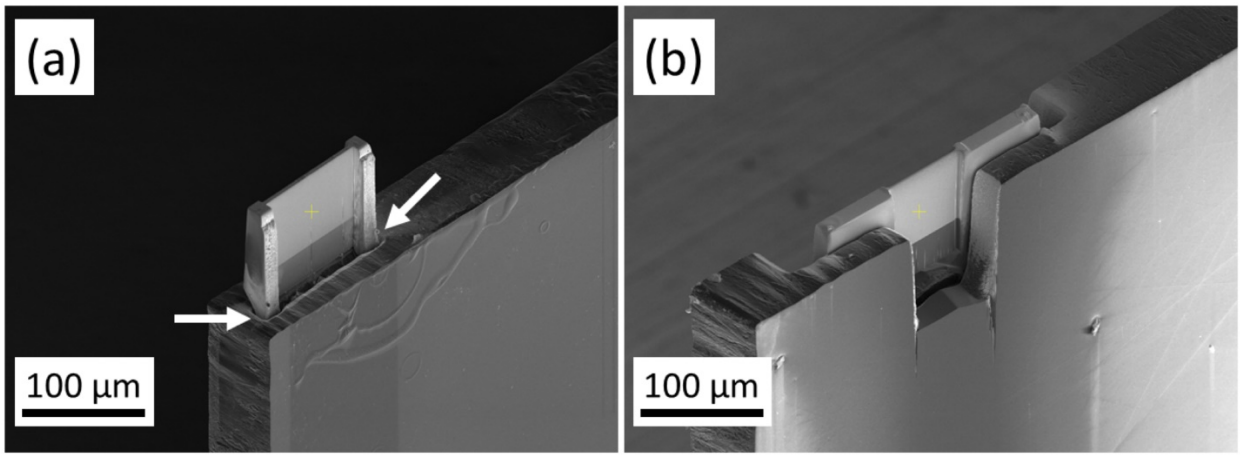

**Fig. 4** (a) SEM image of one of the MLLs used in the XFEL experiment. The MLL is attached to the top of a diamond substrate and has a direct contact to the substrate only via two small Pt dots in the positions indicated by two arrows. (b) SEM image of an alternative geometry where the MLL has much larger contact to the diamond substrate.

We demonstrated that the MLLs can focus, image and survive in XFEL beam. However, current mounting of the MLLs to the substrate has to be improved to enable faster heat dissipation. Hence, we are pursuing new, more robust ways of mounting MLLs, which would keep them stable and in the Bragg condition even at higher beam intensities. One of our new mounting concepts is shown in Fig. 4(b). The top edge of the 100 μm thick diamond wafer was shaped using laser machining and a plasma FIB (PFIB) to accommodate the MLL. The lateral dimensions of the MLL piece were substantially increased as compared to the old design [Fig. 4(a)]. Thus, a much larger area of the MLL can be in direct contact with the diamond substrate, which was additionally polished with PFIB. The edges of the MLL were glued to the substrate with an SEM electron curable glue. These parts are outside the region that would ever be illuminated with an XFEL beam. Only the central part of the MLL (about 100 um$^2$) was thinned to the optimum thickness needed to achieve high diffraction efficiency. The diamond mount was U-shaped so that the thinned part of the MLL exposed to the intense XFEL beam is free standing. This new geometry should be more efficient in transferring heat from the lens to the diamond substrate.

*Disclosures*

The authors have no relevant financial interests in the manuscript and no other potential conflicts of interest to disclose.

*Acknowledgments*

We thank Sabrina Bolmer, Lars Gumprecht, Tjark Delmas, Julia Maracke (CFEL), Shiwani Sharma (The Hamburg Centre for Ultrafast Imaging), Huijong Han and Joachim Schulz (European XFEL) for technical assistance and help with experimental setup, Thomas Keller and Andreas Stierle (DESY) for access to FIB in Nanolab and Andrew Aquila (SLAC), Harald Sinn (European XFEL), Sang-Kil Son, Robin Santra (CFEL) and Ozgur Culfa (Univ. of Nebraska-Lincoln) for helpful discussions. David A. Reis was supported by the AMOS program within the Chemical Sciences, Geosciences, and Biosciences Division of Basic Energy Sciences, US Department of Energy. This work was funded by DESY (Hamburg, Germany), a member of the Helmholtz Association HGF. Additional support was provided by the Deutsche Forschungsgemeinschaft